\documentclass[%
 bibnotes,
 amsmath,amssymb,
 aps,twocolumn,
prb,
]{revtex4}

\usepackage{graphicx}
\usepackage{dcolumn}
\usepackage{bm}

\begin{document}

\title{The Observed Vibron Shift in Helium-Hydrogen Mixtures are due to Quantum Nuclear and Localisation Effects, not bonding}

\author{Sam B. Ramsey}
 \author{Miriam Pena-Alvarez}
\author{Graeme Ackland}%
\affiliation{%
 The University of Edinburgh
}%

\date{\today}

\begin{abstract}

The vibrational frequency of hydrogen molecules has been observed to
increase strongly with He concentration in helium hydrogen fluid
mixtures.  This has been associated with He-H interactions, either
directly through chemical bonding\cite{PhysRevLett.120.165301}, or indirectly through increased local pressure\cite{PhysRevB.32.7611}.
Here, we demonstrate that the increase in the Raman frequency of the
hydrogen molecule vibron is due to the number of H$_2$ molecules participating in the mode.
There is no chemical bonding between He and H$_2$,  helium acts only to separate the molecules.  The variety of possible environments for H$_2$ gives rise to many Raman active modes, which causes broadening the vibron band. As the Raman active modes
tend to be the lower frequency vibrons,
these effects work together to produce the majority of the shift seen in experiment. We used Density Functional Theory (DFT) methods in both solid and fluid phases to demonstrate this effect.  DFT also reveals that the pressure in these H$_@$-He mixture is primarily due to quantum nuclear effects, again the weak chemical bonding makes it a secondary effect. 
\end{abstract}

\pacs{Valid PACS appear here}
\maketitle


Hydrogen and helium are the simplest and most abundant elements in the
universe. As such the recent claim that there is chemical bonding between hydrogen and helium is potentially transformative to understanding their high pressure interactions for  both the condensed matter and astrophysical communities\cite{PhysRevLett.120.165301}. 
The lightness of each element means that nuclear motion and zero-point effects play 
a large part in their dynamics, so that standard methods of electronic structure calculation are insufficient to describe them. This gives rise to exotic phases of matter such as superfluids and, potentially, supersolids\cite{prokof2005supersolid,kim2004probable}. Understanding mixtures of hydrogen and helium under pressure
is important for the study of the gas giants such as Jupiter and
Saturn as they are the primary
constituents\cite{GUILLOT19991183,morales2012rev,Morales03022009}. It is
also important to characterise the mixtures as helium is commonly used
as a pressure medium in diamond anvil cell (DAC)
experiments\cite{DAC}.

Helium and molecular hydrogen are readily miscible in the fluid regime. DAC
experiments $<$10 GPa see a single fluid phase with the characteristic
signal being the Raman active mode of the hydrogen
vibron\cite{PhysRevB.36.3723}.  The vibron frequency is seen to be
blueshifted in mixtures, with the magnitude of this shift being
dependent on the relative concentration of the mixture. This has been variously
attributed to an effective increase of pressure induced on the
hydrogen molecule due to the helium
solution\cite{PhysRevB.36.3723,PhysRevB.32.7611} although no amount of pressure 
can cause such a large shift in pure hydrogen, and to novel chemical bonding\cite{PhysRevLett.120.165301}.
At higher pressures, first H$_2$ and then He solidify into hexagonal close-packed solids and demix, perhaps causing "He-rain" (or more properly, snow) in planetary atmospheres.  Weak
H$_2$ vibrons have been observed in the He-rich solid\cite{PhysRevLett.120.165301}, suggesting low solid solubility.

A clue to the cause of this shift was obtained when a similar effect
was seen in hydrogen mixtures with other elements. The suggested
explanation is that the coupling between hydrogen molecules is
weakened as they are separated by the other elements in
mixtures\cite{PhysRevB.45.12844,PhysRevB.53.R14705,Somayazulu2009}. A simple
classical molecular potential with nearest neighbour interactions has
shown that this effect is of the right order of magnitude to
explain the behaviour in Argon-hydrogen
mixtures\cite{PhysRevB.33.2749,PhysRevB.45.12844}. To our knowledge no
theoretical work has addressed this effect in helium-hydrogen mixtures. Here
we present a first principles investigation of this effect to
accurately describe the observed experimental effects.

To study the system, density functional theory calculations
were carried out on mixtures of helium and hydrogen at
various concentrations. Previous work has concentrated on astrophysically relevant conditions, $~<$100 GPa and $~<$1000 K, where van der Waals interactions and nuclear quantum effects can be safely ignored\cite{mcmahon2012rev,morales2009phase,nettelmann2008ab}.  Our calculations are at relatively low
pressures and require van der Waals interactions, which are accounted for through DFT+D using a PBE+G06
functional\cite{PhysRevLett.77.3865,JCC:JCC20495,doi:10.1063/1.4754130}.
Moreover, below 5GPa, the largest contribution to the pressure comes from quantum nuclear effects: the pressure arising from changes in zero point energy (ZPE) with density. To account for this we carry out standard Nose-Parrinello-Rahman (NPT) calculations, as implemented in CASTEP\cite{0953-8984-14-11-301}, then use lattice dynamics and the quasiharmonic approximation to calculate the true pressure. 
We used molecular dynamics  calculation to model the fluid state, and geometry optimizations on an hcp lattice\footnote{the c/a ratio was taken from the preliminary NPT calculation}. Hydrogen molecules and He atoms were randomly distributed to produce various concentrations.  

 Several thousand molecules are required to fully describe the liquid structure \footnote{H.Geng, M.Marques and G.J.Ackland, submitted}, but the phonon density of states is well sampled in much smaller cells.
 Density functional perturbation theory (DFPT) on small 36 molecule cells were used to calculate Raman activity and the vibrational contribution to the pressure using relaxed lattice and relaxed snapshots taken from the fluid MD after 1 ps equilibration. The DFPT calculations carried out on the resulting metastable state.  The enthalpy was converged to 1 meV  with  using a 2x2x2 kpoint grid. Van der Waals
functionals are essential for helium, however, it is also well known that these functionals overestimate the hydrogen vibron frequency\cite{azadi2017role}. To facilitate comparison with the experimental,  calculated frequencies are shifted by 126 cm$^{-1}$ to match the experimental hydrogen vibron.
All calculations are carried out using the CASTEP code\cite{0953-8984-14-11-301}.

\begin{figure}
\includegraphics[width=1\linewidth]{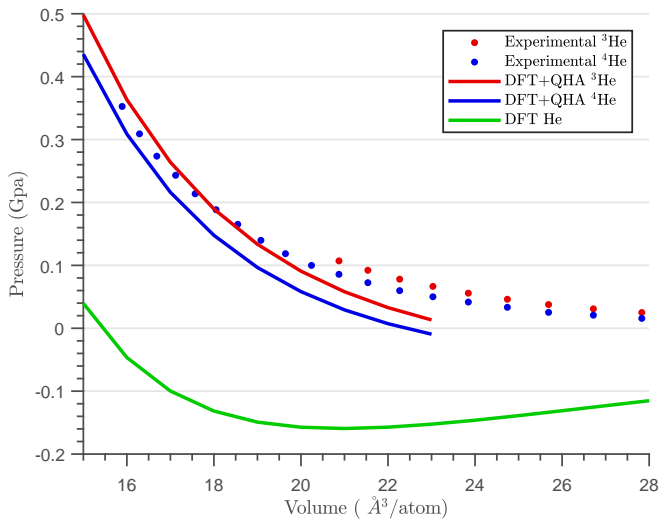}
\caption{Equation of state of both helium isotopes in the
  hcp structure. The DFT  calculations use Grimme van der Waals corrections, for which hcp is stable  against bcc and fcc.  The green line shows the equation of state in the Born-Oppenheimer approximation. Blue and red curves show the effect of adding zero-point pressures in the quasiharmonic approximation.  Above 20 $\AA^3$/atom the static relaxations give negative pressure, and above 23 $\AA^3$/atom the hcp structure cannot be stabilized without ZPE. 
  Experimental data are from X-ray and strain gauge measurements\cite{mao1988high,straty1968compressibility,mills1980equation,PhysRevA.5.2651}. }
\label{fig:EoS}.
\end{figure}

No previous study of solid helium using standard DFT has been published, and Fig. \ref{fig:EoS} suggests why. Van der Waals effects are important, but the dominant contribution to the pressure comes from ZPEs. These have a massive effect on the equation of state, shifting the equilibrium density in pure He by about 50\%.    

\begin{figure}
\includegraphics[width=1\linewidth]{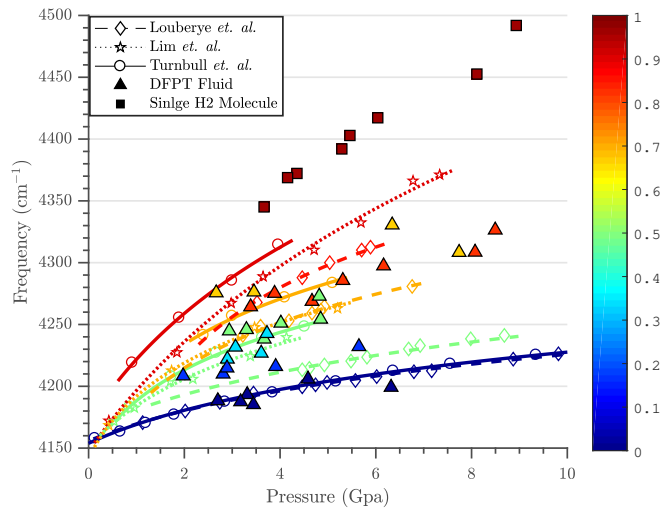}
\caption{Comparison of experimental and calculated pressure-frequency relation with color-coding used to show different concentrations.
 The color scale is represents the concentration of helium: only data of the same color should be compared. Experimental data taken from Turnbull {\it et. al.}\cite{PhysRevLett.121.195702}, Lim {\it et. al.}\cite{PhysRevLett.120.165301}, and Loubeyre {\it et. al.}\cite{PhysRevB.36.3723} is plotted against fluid DFT results for a range of concentrations and Pressures. All experimental data are fitted with a logarithmic function. A a more direct comparison of concentration and frequency is given in Fig. \ref{dos}
\label{FreqPres}}
\end{figure}

The effect of increasing He concentration in these calculations is to blueshift the vibron and to reduce the ZPE pressure. In Fig.\ref{FreqPres}, the calculated variation in Raman shift for different concentrations at different pressures are presented together with experimental results\cite{PhysRevLett.121.195702,PhysRevLett.120.165301}. The  frequencies changes are in good agreement with the experimental results. Analysis of the vibron eigenmodes shows negligible
participation of the He atoms in the motion, demonstrating the absence
of chemical interactions between helium and hydrogen. This suggests the observed {\it blueshift} with increasing helium concentration is due to  fewer couplings between adjacent H$_2$ molecules and localisation of the vibrational modes.

To determine miscibility within the solid regime we examined
H$_2$ impurities in helium.  Both hydrogen and helium atoms form
hcp solid phases in this pressure
regime, and the enthalpy of solution means hydrogen can only occupy only substitutional sites
in the solid helium lattice (Table \ref{tab1}).  At room temperature,
the calculated solubility limit is 0.2at.\%

\begin{table}[]
\begin{tabular}{l|c|c|c}
 & $\Delta$H (eV) & $\Delta$E (eV)  & Miscibility     \\
  \hline
 Substitutional & 0.164 & 0.048 & 0.2\% \\
 Tetrahedral & 0.552 & 1.267 & 2x10$^{-7}\%$ \\
 Octahedral & 0.640 &  1.032 & 5x10$^{-6}\%$ 

 \end{tabular}
 \caption{The enthalpy and energy cost of including hydrogen atoms in hcp helium
   at 12 Gpa for each site is given. The miscibility ($e^{-\Delta H/kT}$) at 300 K is calculated assuming a dilute regular solid solution.}\label{tab1}
\end{table}

Table II shows the binding energy for clusters of substitutional hydrogen
molecules compared with isolated hydrogen molecules in solid helium. All the enthalpies of formation are negative, and becone larger as more hydrogren is added, a strongly suggestion that when within a helium lattice, hydrogen molecules tend to cluster as they attract one ahother. Standard DFT calculations 
suggest strong H$_2$-H$_2$
interactions, relative to He-He,  but unexpectedly this difference is significantly reduced
when the ZPE is accounted for through DFPT.  Nevertheless, below room
temperature the binding is close to the configurational entropy cost, so significant numbers of H$_2$ microclusters can be expected.  The vibrons associated with H$_2$ solutes are significantly blueshifted from the pure H$_2$ value due to the lack of coupling, with the single
substitutional having the largest shift, perhaps accounting for the multiple Raman peaks attributed to interstitials by Yoo {\it et al.}\cite{PhysRevLett.120.165301}.

\begin{table}[h]
\begin{tabular}{l|c|c|c}

 & $\Delta$E  &  $\Delta$E+ZPE(eV) & $k_BT\ln N$  \\
  \hline

 Pair & -0.043 & -0.006 & 0.017 \\
 Triplet & -0.114 &  -0.023 & 0.025 \\
 Quadruplet & -0.185 &   -0.037 & 0.034 \\
 \end{tabular}
  \caption{Formation energy for clusters of substitutional hydrogen
    molecules compared with a lone hydrogen molecule in an hcp helium lattice. Results with and without accounting for zero-point energy are shown, along with the configurational
    entropy cost to the free energy at room temperature.}\label{tab2}
\end{table}

Recent experimental results have claimed that at room temperature and pressures above 12 GPa helium and hydrogen are able to chemically bond\cite{wong2000prediction,PhysRevLett.120.165301}. The experimental evidence for chemical association comes from another vibrational mode observed at around 2400 cm$^{-1}$. We do not see any evidence of any vibrational modes in this frequency regime in our calculations. Turnbull {\it et. al.}\cite{PhysRevLett.121.195702} have demonstrated that this effect can be due to nitrogen.    If, as we propose, the concentration dependence is due simply to the number of hydrogen's participating in the modes, then the effect will be largely independent of whether the hydrogen is diluted by He or N$_2$: by contrast whereas He-H$_2$ chemical bonding or local stresses\cite{PhysRevB.32.7611} will depend on the composition of the solvent. The experiments show consistent vibron shifts with H$_@$ concentration, regardless of nitrogen content.

\begin{figure}%

\includegraphics[width=1\linewidth]{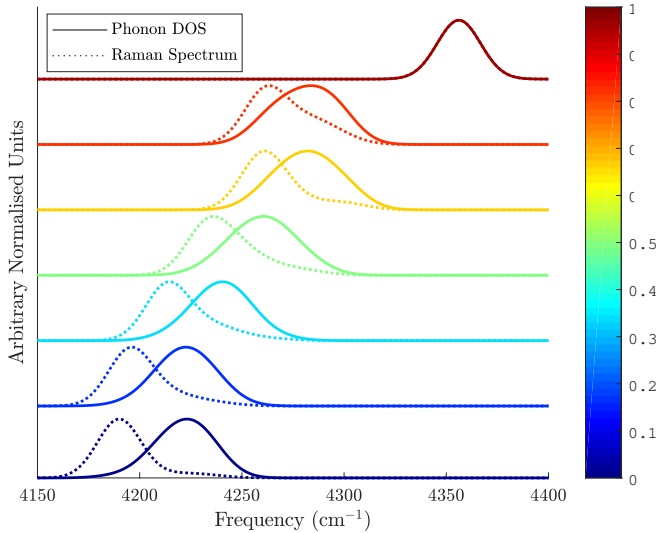}%

\caption{Phonon density of states and associated Raman intensities for a range of concentrations is taken from DFPT calculations carried out in the fluid model showing how the hydrogen vibron mode changes with  concentration (colorbar shows He concentration). The peak broadens and  the average frequency blueshifts shifts  as the He concentration is increased. As the Raman active mode is the lowest frequency of the phonon band this results in the Raman intensities diverging from the phonon density of states at higher H$_{2}$ concentration. All peaks fitted with a 25 cm$^{-1}$ FWHM Gaussian broadening.}
\label{dos}%
\end{figure}

 \begin{figure}
\includegraphics[width=1\linewidth]{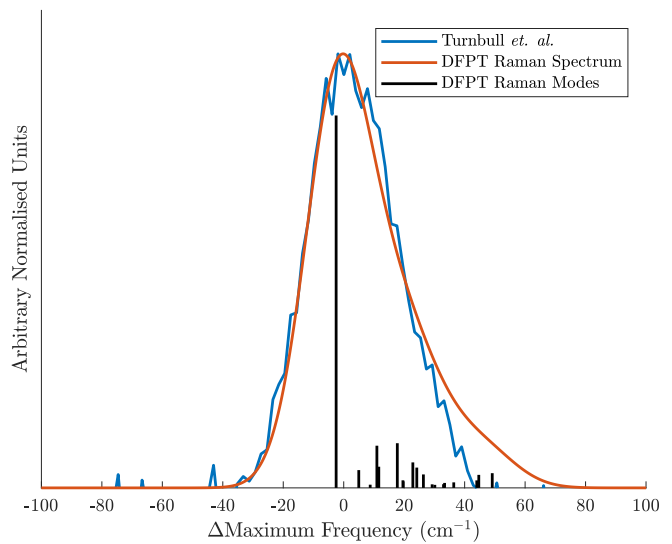}

\caption{Comparison of experimental Raman peak with DFT data at 2.7 GPa and 50\% H$_{2}$ concentration. The DFPT Raman intensities shown in black are convolved with a 25 cm$^{-1}$ Gaussian broadening to get the resulting spectrum. Blue experimental data were collected during a previous campaign  \cite{PhysRevLett.121.195702}.  Both the experimental and DFT spectra are asymmetric with a high frequency tail. This effect is due to the weaker, but still active, Raman modes at higher frequency.}
\label{ramanExp}
\end{figure}

\begin{figure}[ht]
\includegraphics[width=1\linewidth]{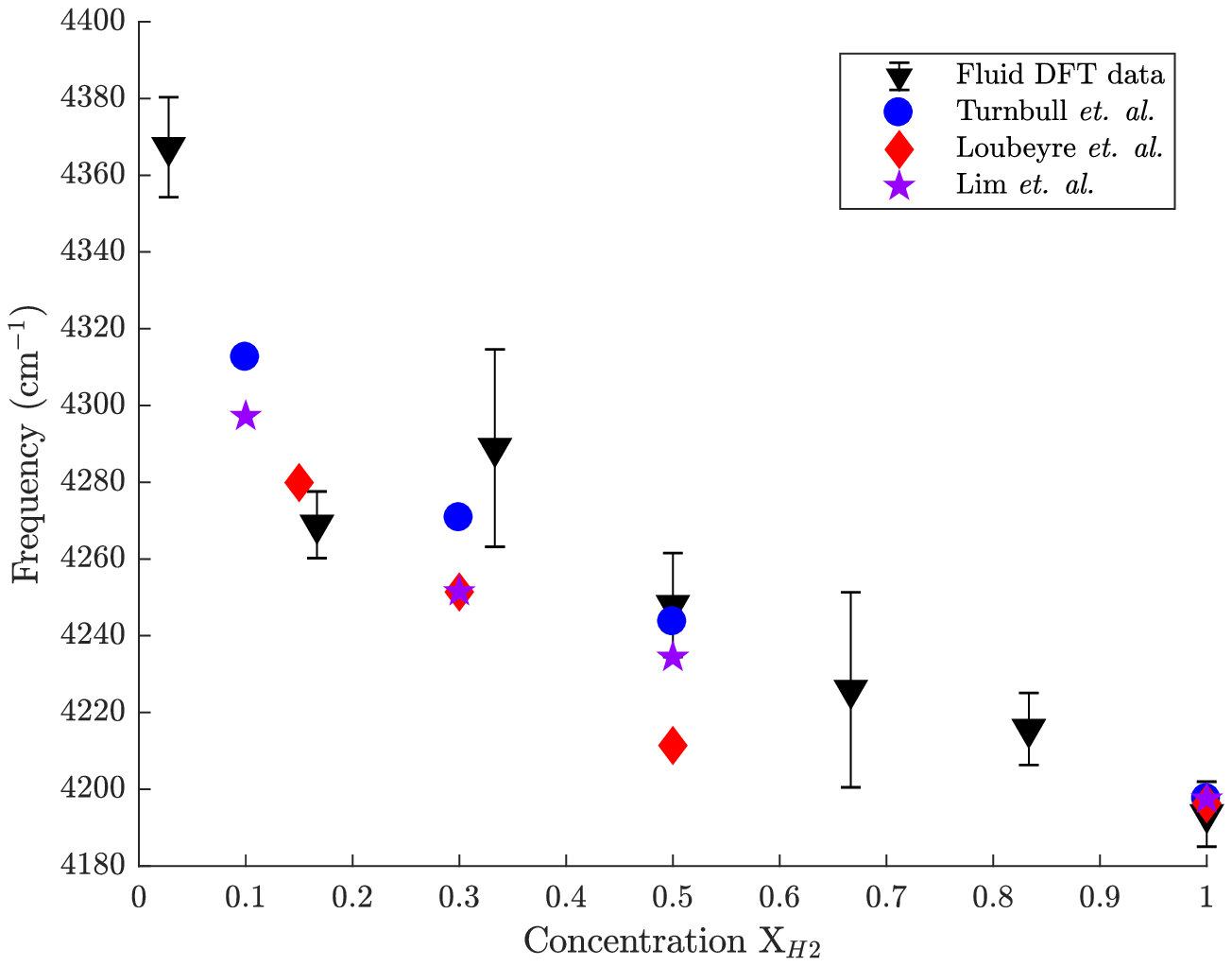}

\includegraphics[width=1\linewidth]{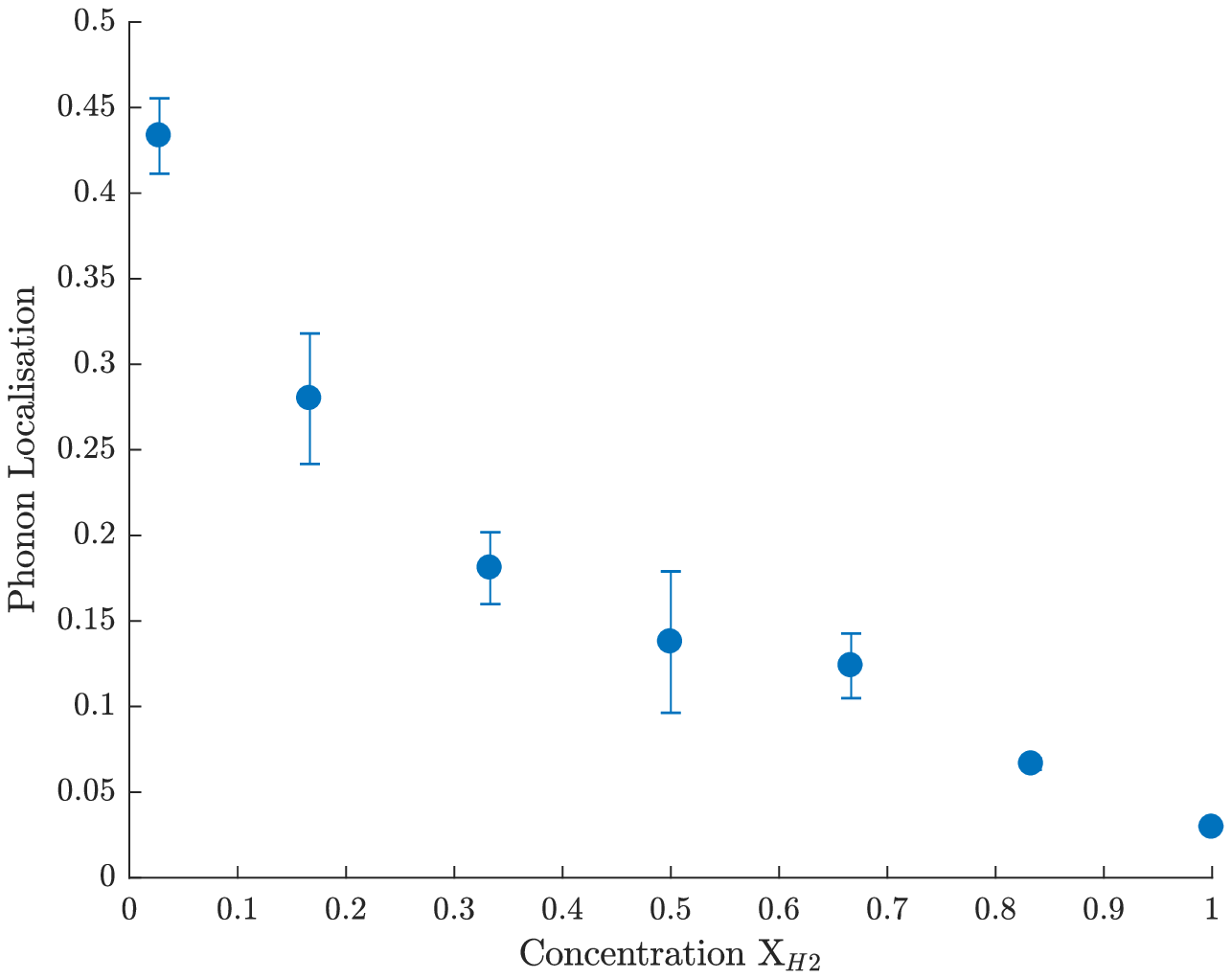}
\caption{(a) DFPT calculation plotted against experimental results\cite{PhysRevLett.121.195702,PhysRevLett.120.165301,PhysRevB.36.3723}, showing the change in Raman active vibron due to change in mixture composition at 4 GPa. DFPT frequencies at precisely 4 GPa are  interpolated from data shown in Fig. \ref{FreqPres}. Error bars on DFPT data are taken from the RMSE of the fits to data in Fig. \ref{FreqPres}. (b) Calculations of the inverse participation ratio of the calculated strongest Raman-active phonon modes.   The error bars are taken from the standard deviation of three independent snapshots at each concentration.}
\label{Local}
\end{figure}

 As shown by Fig.\ref{FreqPres}, an isolated hydrogen molecule, (represented by the X$_{H2}$ = 0.0278), in a helium mixture has a significantly higher Raman frequency than  pure hydrogen at the same pressure. Fig. \ref{dos}a shows that as X$_{H2}$  concentration increases the vibron band is both \textit{broadened} and \textit{shifted}.  
 The apparent phonon frequency shift due to concentration is enhanced by the broadening (Fig\ref{Local}), because  Raman activity tends to be
 stronger for the lower frequency vibron modes. DFPT calculations in the fluid reveals that even the Raman the peak arises from several modes.  The in-phase vibration of all molecules in a cluster is the strongest, but in the absence of symmetry many other modes acquire some Raman activity.  These modes have slightly higher frequency than the in-phase vibron, so cause a skew in the peak shape. \ref{ramanExp}

 Comparison of the panels in Fig.\ref{Local} provides strong evidence that the frequency shift is due to localization, which we measure as the inverse participation ratio:
 \[\frac{\sum_i{\bf e}_i^4}{(\sum_i{\bf e}_i^2)^2}\]
 where ${\bf e}_i$ is the mode displacement vector of each atom $i$.
 The larger the number of hydrogen atoms participating in a vibron mode, the lower its frequency. An isolated hydrogen has the highest frequency, while the lowest observed mode occur in pure solid hydrogen, where the Raman mode involves all molecules vibrating together.  
 The shift and multiplicity of the peaks due to species and fluid disorder are similar to the isotope disorder effect in hydrogen \cite{howie2014phonon,magduau2017infrared}.
 
 These effects are seen in both the solid and the fluid, However, the solid vibron shift is larger because the pure H$_2$ fluid  vibron is already partially localised due to the disordered nature of the fluid. At high H$_2$ concentrations the differences between solid and fluid are largest: all hydrogens have many coupled neighbours, but the fluid vibron is still localised due to the disorder. 

The comparison with the experimental fluid measurements shows that the model
captures the main effects involved in the frequency shift.  The
primary experimental evidence of solid phase miscibility is
observation of an H$_2$ vibron mode in the
helium\cite{PhysRevLett.120.165301}.  Consistent with our calculations
for {\it substitutional} and {\it clustered} H$_2$, these observed
modes are blueshifted with the less-blueshifted cluster mode being
broader. It was not possible to determine a precise H$_2$
concentration in the experiment\footnote{C.S. Yoo, private communication}, but our
calculated solubility is sufficient to produce an observable signal
and therefore the calculations support the experimental data\cite{PhysRevLett.120.165301}.

 In our DFPT calculations we have assumed that two elements are
 randomly distributed throughout the mixtures. However, if the
 hydrogen molecules are more clustered this would enhance coupling and
 drive down the vibron frequency of the low X$_{H2}$ mixtures. To
 understand the potential magnitude of this effect DFPT calculations
 where carried out on a simulation cell with X$_{H2}$ = 0.1667 with a single cluster of
 hydrogen molecules:  This resulted in a drop in frequency of 22 cm$^{-1}$, and reduced inverse participation ratio. Thus we show that localization increases the frequency, independent of concentration.

Our results can be used as a future reference for experimental works on mixtures and alloys in general, and on hydrides in particular. Experimental mixtures are sometimes prepared {\it{in situ}} and the real concentration of the components may not be known. However, here we show that a relationship between Raman shift, pressure and hydrogen concentration which could be used as a reference concentration calibrations. On the other hand, this work highlights that attention should be paid not only to the shifting of the vibrons but also to the width as it can provide important information about concentrations and inter molecular interactions.

In conclusion we have carried out the first ab-initio study of the
hydrogen molecular vibron in HeH$_2$ mixtures. We show that including van de Waals corrections and zero point energy effects are essential to reproducing the equation of state below 10 GPa.
The vibron blueshift with increasing He concentration is shown
to be due to the reduction of hydrogen-hydrogen coupling and the associated
localisation of the vibrational mode.  "local pressure" effects can be ruled out because the isolated hydrogen molecule
in He has a significantly higher vibron frequency than pure hydrogen
at any pressure.  The observed broadening of the vibron in mixtures is because there are Raman active vibrations involving various 
numbers of H$_2$ atoms.  The calculations support the possibility of small amounts of H$_2$ existing as substitutional impurities in solid He, but unequivocally rule out interstitial H$_2$ or any He-H$_2$ chemical bond.

\begin{acknowledgments} We acknowledge E.Gregoryanz and R.Howie for useful discussions and for sharing their primary data, and C.S.Yoo regarding the observability of low H$_2$ concentrations in the He-rich fluid.  GJA and MPA were supported by the ERC Hecate project, and SBR acknowledges a studentship from EPSRC.  Computing resources were provided by UKCP (EPSRC grant EP/P002790)
\end{acknowledgments}
\bibliography{Bibliography}       

\end{document}